\documentclass[twocolumn,secnumarabic,amssymb, superscriptaddress, nobibnotes, aps, prl]{revtex4-2}

\usepackage{graphicx}
\usepackage{color}

\begin{document}


\title{Controlled creation and annihilation of stringless robust emergent magnetic monopoles in artificial spin ice}
\author{N. Keswani}
\affiliation{Department of Physics, Indian Institute of Technology, Delhi, New Delhi 110016, India}
\author{R. Lopes}
\affiliation{Departmento de Fisica - Universidade Federal de Vicosa, Brazil}
\author{Y. Nakajima}
\affiliation{Bio-Nano Electronics Research Centre, Toyo University, Saitama 3508585, Japan}
\author{R. Singh}
\affiliation{Institute of Physics, Sachivalaya Marg, Bhubaneswar 751005, Odisha, India}
\author{N. Chauhan}
\affiliation{Bio-Nano Electronics Research Centre, Toyo University, Saitama 3508585, Japan}
\author{T. Som}
\affiliation{Institute of Physics, Sachivalaya Marg, Bhubaneswar 751005, Odisha, India}
\author{S. Kumar}
\affiliation{Bio-Nano Electronics Research Centre, Toyo University, Saitama 3508585, Japan}
\author{A. Pereira}
\affiliation{Departmento de Fisica - Universidade Federal de Vicosa, Brazil}
\author{P. Das}
\email{pintu@physics.iitd.ac.in}
\affiliation{Department of Physics, Indian Institute of Technology, Delhi, New Delhi 110016, India}

\date{13/11/2020}%

\begin{abstract}

Magnetic analogue of an isolated free electric charge, i.e., a magnet with a single north or south pole, is a long sought-after particle which remains elusive so far. In magnetically frustrated pyrochlore solids, a classical analogue of monopole was observed as a result of excitation of spin ice vertices. Direct visualization of such excitations were proposed and later confirmed in analogous artificial spin ice (ASI) systems of square as well as Kagome geometries. However, such charged vertices are randomly created as they are thermally driven and are always associated with corresponding emergent antimonopoles of equal and opposite charges connected by observable strings. Here, we demonstrate a controlled stabilisation of a robust isolated emergent monopole state in individual square ASI vertices by application of an external magnetic field. The excitation conserves the magnetic charge without the involvement of a corresponding antimonopole. Well supported by Monte Carlo simulations our experimental results enable,  in absence of a true elemental magnetic monopole, creation of electron vortices and studying electrodynamics in presence of a monopole field in a solid state environment.


\end{abstract}
\maketitle


Following the seminal theoretical work of Dirac~\cite{Dirac1931, Dirac1948} predicting the existence of a magnetic monopole, search of monopoles has been a major theme in physics~\cite{Rajantie2017, Qi2009, Uri2020, Goldhaber1990}. The theory shows that the existence of monopole is a precondition for quantization of electronic charge ($e$) which is given by $C_m e=\frac{1}{2}hc$, where $C_m$, $h$ and $c$ are the strength of a magnetic pole, Planck's constant and vel. of light, respectively~\cite{Dirac1931, Dirac1948}. Therefore, the quest for finding signature of stable magnetic monopoles led researchers to investigate in vastly different platforms such as from cosmic radiation to accelerator based high-energy experiments~\cite{Milton2006, Rajantie2017}. On a different platform, Hall \textit{et al.}, created a nonmagnetic monopole-like state in optical traps using ultra cold rubidium atoms in a Bose Einstein Condesate state~\cite{Ray2014}. A classical analogue of a monopole was predicted~\cite{Castelnovo2008} and experimentally observed~\cite{Jaubert2009, Morris2009} in tetrahedral pyrochlore solids where the Ising-like large $f$-electron spins of rare-earth ions interact following a local principle of 2-spins in/2-spins out of  tetrahedra, called spin ice rule~\cite{Harris1997}. There, the monopole is created as a collective behavior of an excitation from a divergence-free low-energy spin ice state which can be considered as vacuum for the local excitation~\cite{Castelnovo2008}. 
\begin{figure}
\includegraphics [width=1\linewidth]{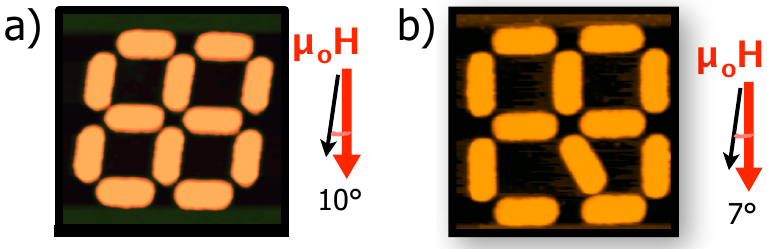}
 \includegraphics [width=0.95\linewidth]{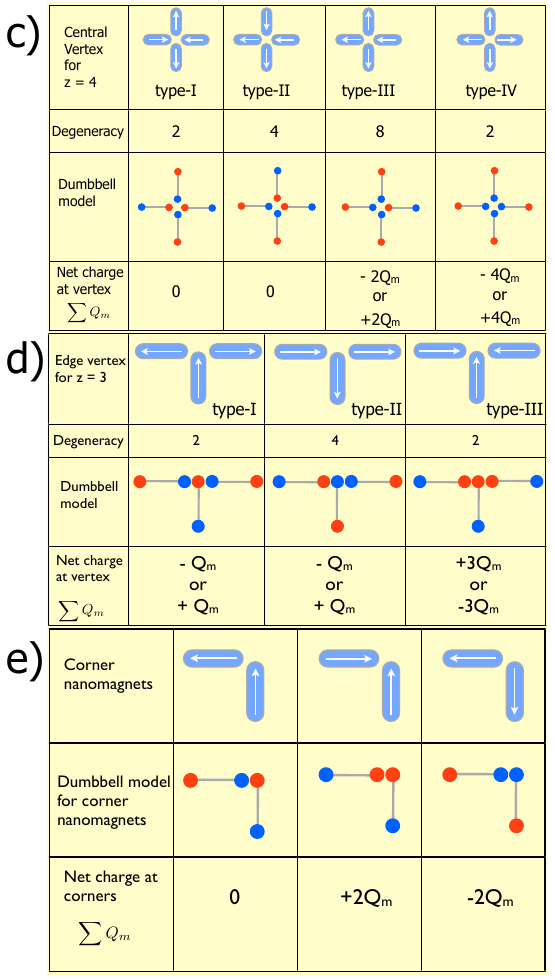}
    \caption{\label{Fig1} {\bf Topography images of ASI systems as well as schematics defining magnetic charges at border and central vertices as well as topography images of ASI systems.}\\ AFM images of square ASI vertices with closed edges resembling a stained glass window {\bf(a)} and a deformed stained glass window {\bf(b)}. The stadium shaped nanomagnets are of dimensions $300\times100\times25$\,nm$^3$ and are in magnetically in single domain state. In the deformed stained glass window, the misaligned nanomagnet is placed at 30$^{\circ}$ with respect to the long axis of the vertical nanomagnets. The applied magnetic field direction is at an angle of $10^{\circ}$ and $7^{\circ}$ with respect to the vertical nanomagnets for the window patterns (a) and (b), respectively. {\bf(c)} shows the possible vertex configurations for $z=4$ vertex with nomenclatures of different types (type-I, type-II,...etc.) of spin ice vertices. Schematics of possible orientations of magnetization, magnetic charges in the nanomagnets under dumbbell model and net charges at edge vertices for $z=3$ {\bf(d)} and corners {\bf(e)}. }
\end{figure}
The emergent monopole in spin ice system carries a magnetic charge and is a non-local entity which is connected to a corresponsing antimonopole by a "string" of spins aligned along one direction thereby maintaining the charge neutrality in the system~\cite{Jaubert2009}. These emergent monopole-antimonopole paired states were observed in 2-dimensional analogue of spin ice~\cite{MorganNatPhys2010, Ladak2010,  Mengotti2010}  where dipolar interactions among nanomagnets of strong shape anisotropy mimicking the Ising spin like behavior leads to spin frustration, which in these cases, is by design~\cite{WangNature2006, NisoliRMP2013, RougemailleEJP2019, Skjaervo2020}. The excitation of such magnetic charges, which appear to interact via Coulomb-type interaction, are typically observed in large arrays of ASI vertices. They are thermally induced and therefore, random in space and time~\cite{Mengotti2010, LadakNJP2011, FarhanScience2019}. To be useful for fundamental studies or practical applications involving a monopole field, two major challenges are encountered in this case: first, controlled creation and annhiliation of such emergent monopole. Second, stabilization of an emergent monopole without a corresponding antimonopole. These are also of profound theoretical interest in general~\cite{Kazama1977}. In this work, we investigated the possibility of controlled stabilization of emergent monopole in isolated square ASI vertices of coordination no. $z=4$ by using external magnetic field as a control parameter. 
Investigation of finite-size ASI systems, particularly the information on how magnetic charges are formed in finite size systems are very limited in literature~\cite{NisoliRMP2013, RougemailleEJP2019, Skjaervo2020, KeswaniJAP2019}. Our work was inspired by a recent report of micromagnetic simulations for finite-size-ASI systems involving elliptical-shaped nanomagnets in ref.~\cite{KeswaniJAP2019}. 

\begin{figure}
\includegraphics [width=1\linewidth]{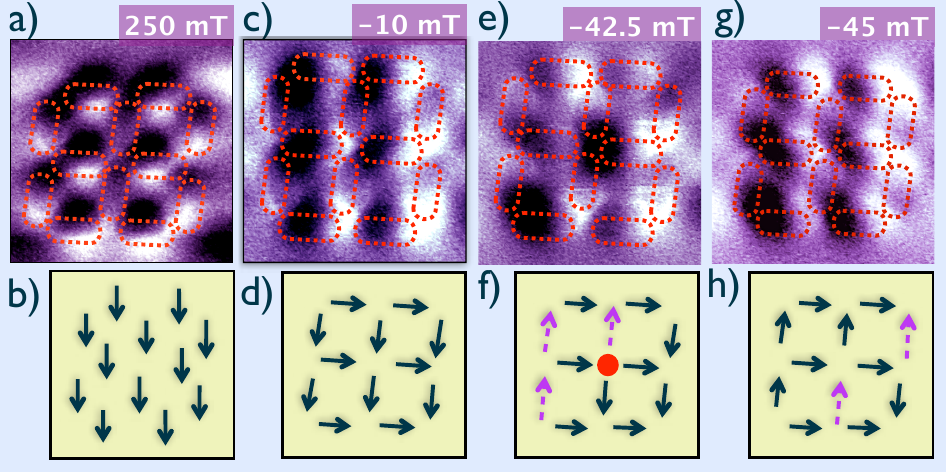}
\includegraphics [width=1\linewidth]{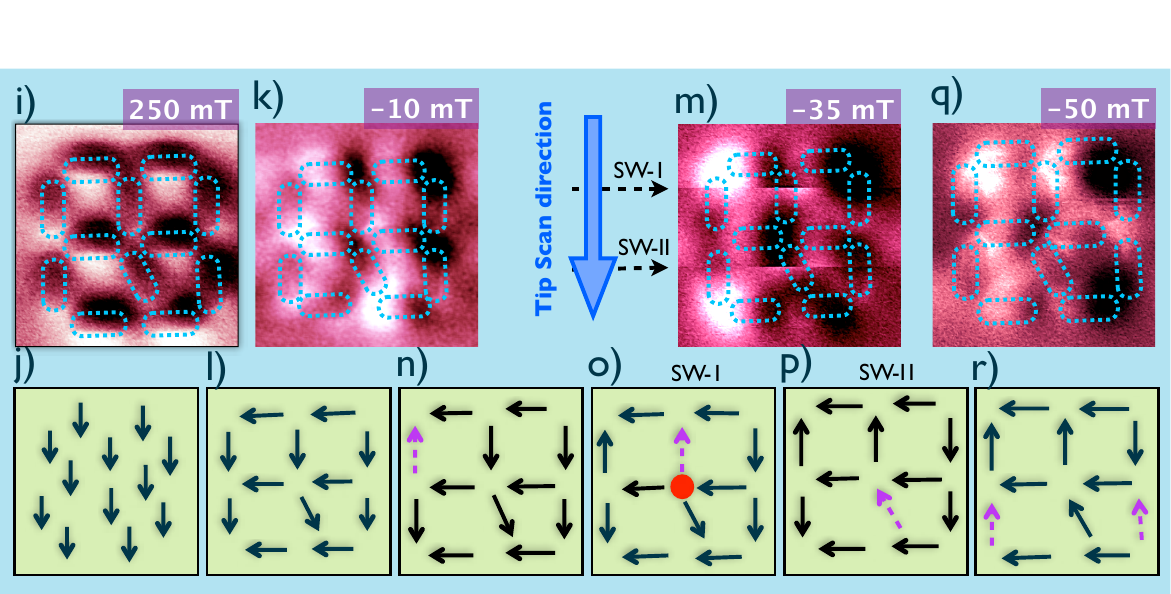}
\includegraphics[width=1\linewidth]{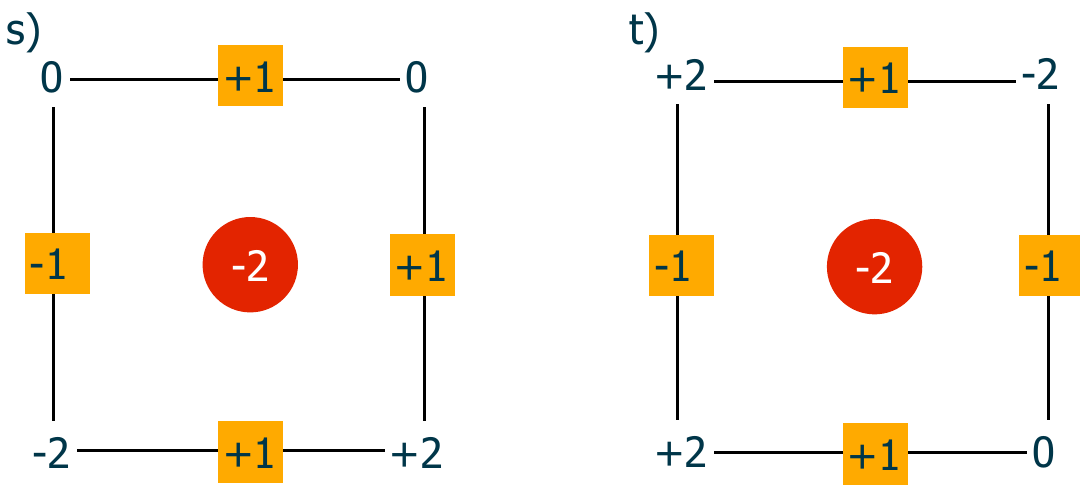}
   \caption{\label {Fig2} {\bf MFM images at discrete magnetic fields and the corresponding arrow diagrams for the two samples (Fig.\,\ref{Fig1}(d,e)) and description of magnetic charges.}  {\bf(a)}, {\bf(c)}, {\bf(e)} and {\bf(g)} are MFM images of the stained glass window sample (Fig.\,\ref{Fig1}d) at $\mu_0H_{\rm{ext}}=$ 250\,mT, -10\,mT, -42.5\,mT and -45\,mT, respectively. {\bf(b)}, {\bf(d)}, {\bf(f)} and {\bf(h)} are corresponding arrow diagrams clarifying the orientation of the net magnetizations in the nanomagnets. The circles at the central vertex position in {\bf(f)} and {\bf(o)} indicate magnetically charged emergent monopole states. {\bf(i)}, {\bf(k)}, {\bf(m)} and {\bf(q)} are MFM images of the deformed stained glass window sample (Fig.\,\ref{Fig1}e) at $\mu_0H_{\rm{ext}}=$250\,mT, -10\,mT, -35\,mT and -50\,mT, respectively. {\bf(j)}, {\bf(l)}, {\bf(n)}-{\bf(p)} and {\bf(r)} clarifies the orientation of magnetization at the corresponding fields. The dotted stadium shapes in the MFM images indentify the actual shapes of the nanomagnets and the dotted (magenta) arrows in the arrow diagrams for both samples show the switched nanomagnets at the corresponding fields. {\bf(n)}-{\bf(p)} show the three switchings at the bias field of -35\,mT. {\bf(o)} and {\bf(p)} are tip-induced switchings (Fig.\,\ref{Fig2}m) at tip scan line positions indicated by SW-I and SW-II while tip scanning in the downward direction. The charge distribution in units of $Q_m$ in the monopole states for undeformed {\bf(s)} and deformed stained glass window {\bf(t)}.}

\end{figure}

We fabricated isolated vertices comprising of strongly shape anisotropic stadium-shaped nanomagnets of Ni$_{80}$Fe$_{20}$ of dimensions 300$\times$100$\times$25\,nm$^3$, in square ASI geometry using electron beam lithpgraphy followed by lift-off process. As shown in Fig.\,\ref{Fig1}a, atomic force microscopy (AFM) image of the sample appears as a stained glass window. The nanomagnets are magnetically in single domain state (see MFM images in Fig.\,\ref{Fig2}) and can be treated as switchable macrospins. We used external magnetic field as a control parameter which was applied along the easy axis of one of the islands in the reduced sub-lattice so that there is a finite probability of preferential switching of one vertex-island. In such a vertex where a magnetic charge can exist only as an excitation, a preferential switching of a vertex-island is necessary for the creation of an excited emergent monopole.

The magnetization state of the system was probed using magnetic force microscopy (MFM) technique (see methods). The applied field direction is along [0 $\bar{1}$], practically which lies at an angle of $\sim\,$10$^\circ$ with respect to the easy
axis of the vertical nanoislands as shown in Fig.\,\ref{Fig1}a. We first polarized the sample at a field of $\mu_0H_{\rm{ext}}=+250$\,mT and then investigated the response of the nanomagnetic stained glass window by collecting magnetic images at discrete field values while reversing the field. The images were collected at every 2.5\,mT so that the complete information of magnetization reversals of the 12 nanomagnets is obtained. Magnetic images collected at saturation (250\,mT), -10\,mT, -42.5\,mT and -45\,mT, are shown in Fig.\,\ref{Fig2}(a, c, e and g), respectively. 

As shown in Fig.\,\ref{Fig2}a, dark and bright patches at the long edges (short edges) of the horizontal (vertical) nanomagnets show that their magnetizations align along the external field direction confirming the saturation at $\mu_0H_{\rm{ext}}=250$\,mT. This is clarified by a corresponding arrow-diagram, where the arrows show the magnetization direction in the nanomagents (Fig.\,\ref{Fig2}b). While reversing the field, we observe at remanence the magnetization of horizontal nanomagnets, for which the external field is along their hard axes, rotate towards their easy axes whereas that of all the vertical nanonanomagnets are still oriented along the saturation field direction thus producing a 2-in/2-out state at the central vertex which is a spin ice state of type-II (see Fig.\,\ref{Fig1}c). An image collected for  $\mu_0H_{\rm{ext}}=-10$\,mT (Fig.\,\ref{Fig2}c) shows the same magnetic state as that of remanence suggesting that no switching has taken place till this field. Considering each dipole of moment $\mu$, which is a switchable macrospin in this case, as a dumbbell of magnetic charges $\pm Q_m$ (=$\mu/d$) separated by a distance $d$, the central vertex with $z=4$ is at a chargeless ($\sum Q=0$) state~\cite{NisoliRMP2013}. Our MFM data for $-10\,\rm{mT} \leqslant \mu_0H_{\rm{ext}}\leqslant -40$\,mT do not show any significant change in the magnetic states of the nanomagnets indicating that no magnetization switching occurs in this field range. At $\mu_0H_{\rm{ext}}=-42.5$\,mT, we observe reversals of bright and dark patches for three vertical nanomagnets suggesting the switchings of their magnetizations (see Figs.\,\ref{Fig2}e, f). These observations suggest that the switchings have taken place within the small field range of -40\,mT$< \mu_0H_{\rm{ext}}\leqslant$-42.5\,mT. Our careful analysis indicates that the three nanomagnets switch simultaneously (see below). Interestingly, these switchings lead to the creation of an excited 3-out/1-in state thereby violating the spin ice rule at the central vertex. This excited state at the vertex, which now has a non-zero magnetic charge $\sum Q=-2Q_m$, is stable against mutliple scanning by the magnetic tip at the given external field. Thus, a stable isolated emergent monopole is created at the central vertex of the stained glass window. 
In general, magnetically charged vertices in arrays of ASI are always found to occur in pairs of equal and opposite charges (pair of monopole and antimonopole) separated by a string of chargeless vertices~\cite{Ladak2010, Mengotti2010, Perrin2016}.
The energy cost of separating such oppositely charged monopole and antimonopole in an array is proportional to the number of chargeless vertices involved between the pair~\cite{Pereira2013, Nisoli2018}. Our experiments reveal that such excited charged magnetic states can be reproducibly stabilised without any corresponding antimonopole. This emergent monopole can be annihilated again by external field. We find that upon increasing the field by 2.5\,mT, i.e., at a field of -45\,mT, the magnetization of two more nanomagnets switch simultaneously (Figs.\,\ref{Fig2}g, h) leading to the vertex converting to chargeless type-II spin-ice state.  
Thus, these results demonstrate the controllability of the creation and annihilation of an emergent monopole state in the form of a magnetically charged vertex which is remarkable given the reported experimental results so far demonstrated the random creation as a paired monopole-antimonopole state~\cite{RougemailleEJP2019, Nisoli2018}. 

\begin{figure}
\includegraphics[width=1\linewidth]{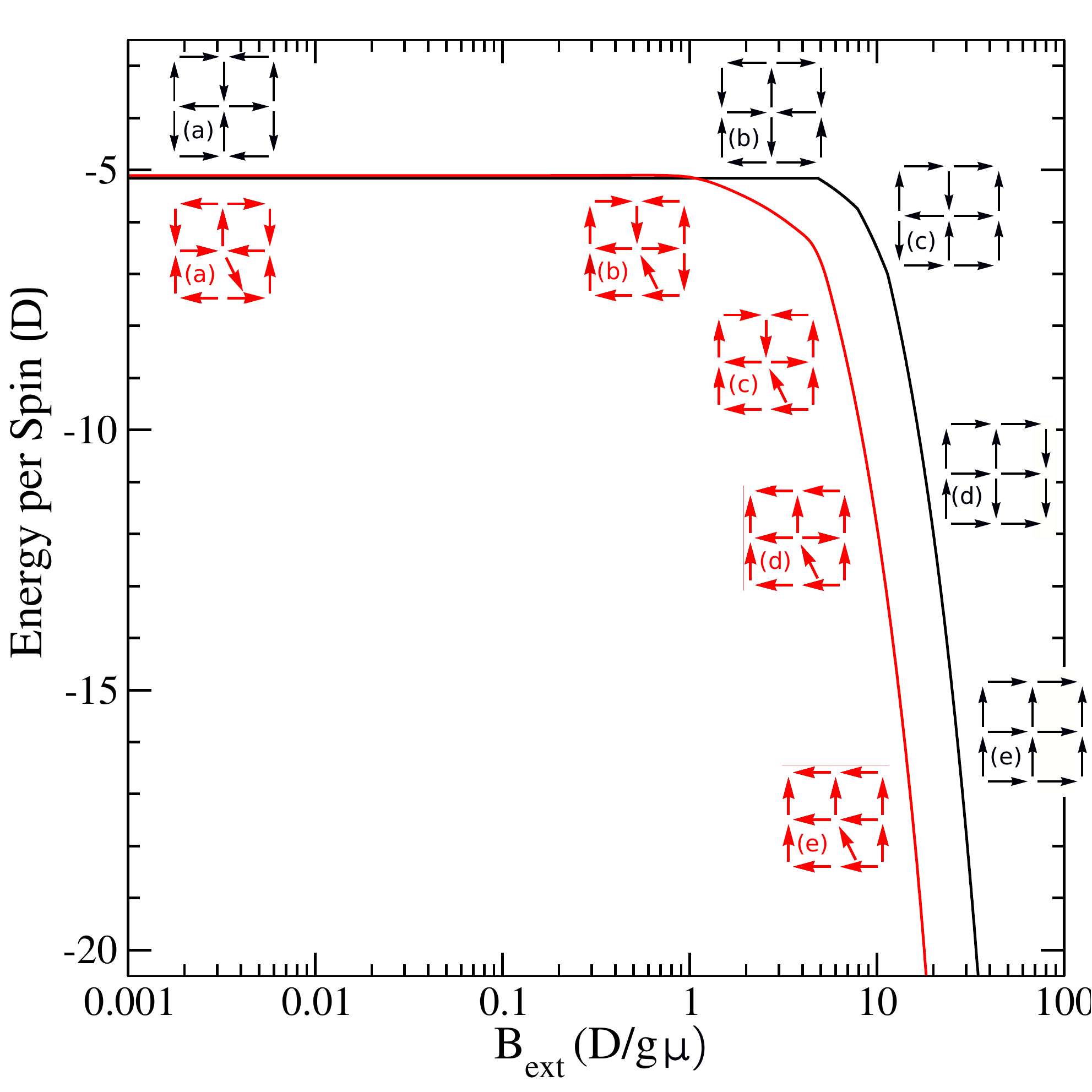}
    \caption{\label{Fig3} \textbf{Most probable, minimum energy magnetic configurations for the undeformed (black) and deformed (red) windows plotted against external field.}  Magnetic field is applied at an angle of 10$^\circ$ with respect to the easy axis of the vertical nanoislands for the former and 7$^\circ$ for the latter sample. Configurations $a$ and $b$ are the possible ground states for both samples. For undeformed samples, as the external field increases to near $5 D/g \mu$, configuration $b$ still shows the most probable state. With further increase of field, configurations $c$, $d$ (with a central monopole) and $e$ (type-II) are observed. The same behavior is observed for the deformed window sample. }
\end{figure}

The controlled creation and annihilation of such monopole offers a possibility to utilize them in designing novel experiments. However, it is essential to determine the robustness of the monopoles in such systems. In order to investigate the robustness of such isolated monopoles against any defect in the vertex, we created an artificial defect with an aim to introduce a perturbation in the dipolar interactions among the vertex-nanomagnets. The defect is in the form of canting of one of the nanomagnets at an angle of $30^{\circ}$ with respect to the vertical nanoislands. Figure\,\ref{Fig1}b shows the AFM image of the defective sample. The canting or misalignment results in a distortion of net local field at the vertex. Except for the misalignment, all other dimensions of the nanomagnets are same as for Fig.\ref{Fig1}a. The patterned defective structure in Fig.\,\ref{Fig1}b, which can be considered as a deformed stained glass window, was placed in an in-plane external magnetic field such that the field was at an angle of $\sim7^{\circ}$ with respect to the vertical nanoislands. This has an intersting consequence in the magnetization switchings as discussed below. The MFM images (Fig.\,\ref{Fig2}\,i,k,m,q) of the deformed window were collected under the same condition and same field protocol as for the defect-free case discussed above. Similar to the case of undeformed window, here also saturation (at $\mu_0H_{\rm{ext}}=\,$250\,mT, Fig.\ref{Fig2}i)  and type-II vertex at remanence are clearly observed. Fig.\,\ref{Fig2}k shows the near-remanence data collected at $\mu_0H_{\rm{ext}}=-\,$10\,mT where no switching of magnetization took place so far. The first switching is observed in the MFM image collected at $\mu_0H_{\rm{ext}}=\,$-35\,mT (Fig.\,\ref{Fig2}m, n). Additionally, two tip induced switchings are observed at the same external field which are due to the magnetostatic interactions between stray fields emanating from the magnetic tip and the respective nanomagnets. The external field induces the first switching which occurs for an edge-nanomagnet exemplified in the arrow diagram in Fig.\,\ref{Fig2}n. We find that the 1st switching field in this case is reduced by about 18\% compared to the defect-free case. Interestingly, this is of the same order by which the net dipolar interaction energy is reduced suggesting that such defects may be used as a tuning parameter for controlling switchings in such strongly dipolar coupled systems. We note here that the magnetic configuration at the vertex after this switching remains in 2-in/2-out spin ice state (Fig.\,\ref{Fig2}n). Further interesing observations are made as the magnetic tip scans over the deformed window. As scanning proceeds (downward in the
image in Fig.\,\ref{Fig2}m), the magnetization of one of the vertex-nanomagnets (see arrow diagram Fig.\,\ref{Fig2}(o)) suddenly switches. The corresponding nanomagnet which showed dark-bright patches at the edges is now observed as dark-dark patches suggetsing that the nanomagnet's magnetization switched almost midway (after dark edge is imaged) during the scan. This tip-induced switching due to the local tip-field ($\sim 20$\,mT) creates an excited 3-out/1-in magnetically charged emergent monopole state at the vertex. As the tip continues to scan further downward and reaches the canted nanomagnet, it prompts  the 3rd switching at the same external field (Fig.\,\ref{Fig2}p) which brings the vertex back to the chargeless spin ice state. These tip-induced switchings at the same external field of -35\,mT, clearly indicate the lowering of the barrier energy for switching due to the introduction of the canted defect. Multiple scans over the sample led to the same result demonstrating excellent reprodicibility of the results. These observations demonstrate that the intrinsic magnetization dynamics of the individual nanomagnets and their mutual interactions at an external field of -35\,mT results in to a metastable magnetically charged emergent monopole state which can be created and annhiliated by the stray field from the cantilever tip. We believe that this monopole can be stabilised in an external field which has been offset by the tip's magnetic field. Indeed, in a separate experiment with a cantilever tip of low magnetic moment, a stable monopole was created at a larger field (not shown).

Thus we find strong experimental evidence for robust isolated emergent monopoles in finite-size $z=4$ square ASI vertices which can be stabilised without any corresponding antimonopole. In typical ASI systems, the monopole-antimonopole pairs of opposite charges maintain the charge neutrality in the system. In order to understand how the charge neutrality is reconciled in the isolated monopole state, we analyze the magnetic charges at the central vertex as well as the edges. The four vertices with $z=3$ at the edges are in either 2-in/1-out or 2-out/1-in state and are naturally charged with absolute value of $\sum Q_{\rm{edge-vertex}}=Q_m$ (see Fig.\,\ref{Fig1}d) whereas the absolute value of net charges at the corners are either $\sum Q_{\rm{corner}}=2Q_m$ or $\sum Q=0$ (Fig.\,\ref{Fig1}e). 3-in/0-out or 3-out/0-in state causing charge of $\pm 3Q_m$ at $z=3$ vertex is energetically unfavorable. Considering the central vertex as well as all the border charges, we find that both the windows exhibit zero net magnetic charge ($Q_{\rm{TOTAL}}$) after each switching. Thus, although the monopole defect is of charge $\sum Q= 2Q_m$, the charges at the boundary  arrange themselves so that $Q_{\rm{TOTAL}}= Q_{\rm{monopole}}+Q_{\rm{boundary}}=0$ suggesting that the net magnetic charge of both the stained glass windows remain conserved at the emergent monopole state even in the absence of a corresponding antimonopole. The total magnetic charge distribution at the monopole state for undeformed window (Fig.\ref{Fig2}f) and deformed window (Fig.\ref{Fig2}(o)) are shown in Figs.\,\ref{Fig2}s, t, respectively.

We next investigated the validity of this remarkable experimental realization of isolated emergent monopole state in stained glass window samples theoretically by performing
Monte Carlo simulations at different external fields ($\vec{B_{\rm{ext}}}$). By considering the exact field configuration as in the experiments for both samples, we calculated the corresponding energy and probability for all possible states from which we determined the most probable magnetic configuration for the two systems at room temperature. As shown in Fig.\,\ref{Fig3} (black curve) where energy per spin is plotted against $B_{\rm{ext}}$, the configuration of the interacting macrospins in the undeformed stained glass evolves as follows: firstly, for very low field ($B_{\rm{ext}} \approx 0$), the degenerate ground state (type-I vertices of configurations $a$ and $b$ in the figure with antiferomagnetic ordering of macrospins) is observed.
However, for $B_{\rm{ext}} \sim1D/g \mu$, the degeneracy is broken as the configuration
$b$ becomes energetically more favorable. Here, $D$ is the dipolar interaction constant, $g$ is the gyromagnetic factor and $\mu$ is the net average magnetic moment of the nanoislands. As the field increases further from $1D/g\mu$, the configuration $c$  is stabilised. Upon further increase of field, a magnetically charged vertex (configuration $d$) becomes energetically most stable for $B_{\rm{ext}}\sim 20D/g\mu$. The charged vertex remains stable till about $B_{\rm{ext}}\sim 25D/g\mu$ before it converts to a chargeless type-II state (configuration $e$). The system then saturates at a high field (not shown). Similar behavior is observed for the deformed stained glass-like sample (red curve in Fig.\,\ref{Fig3} ) where the field is applied at an angle of 7$^{\circ}$ in order to match the experimental conditions. Interestingly we find that the corresponding configurations for this case occur at reduced energies which is also observed in the experiments as discussed above. The emergent monopole state for this sample is observed at a relatively lower field  $B_{\rm{ext}}\sim10D/g\mu$ which again changes to type-II chargeless vertex state at $B_{\rm{ext}} \sim15D/g\mu$ (Fig.\,\ref{Fig3}). Quantitatively, considering the average magnetic moment of each nanoisland $\mu\sim9.65\times10^{-12}$ emu , we find dipolar interaction energy, $D\approx3.86\times10^{-17}$J for the undeformed samples whereas $D\approx6.76\times10^{-17}$J for the deformed sample. For the calculations, the monopole fields of 40\,mT and 35\,mT, respectively were considered for the undeformed and deformed samples, respectively. Thus, our experimental findings of emergent monopole state in conjunction with Monte Carlo simulations provide an estimation of the dipolar interaction energy. 

The analysis of magnetic charges observed in our Monte-Carlo simulations demonstrate a clear resemblance with the experimental results underlining the observation that the net magnetic charge is conserved at the emergent monopole state. The monopole like excitations at the centre of the vertices are screened by the magnetic charges at the borders. Thus, we find a clear evidence that the conservation of magnetic charges is a precondition for stabilization of isolated monopole state in square ASI vertices. We note here that any change in magnetic charge at the boundary (surface) annihilates the corresponding monopole at the vertex (in bulk). Thus, our experimental observations of the stabilization of isolated emergent monopoles are in excellent agreement with the Monte Carlo based calculations which delineate a clear condition for creating such excited states in finite-size ASI systems. The results also suggest that the creation of paired emergent monopole-antimonopole pair is not a necessary requirement for preserving the charge neutrality in a large array of ASI vertices. The neutrality may be locally conserved through the border charges as we observe in our studies. This precondition implies that the multipole expansion for the potential of the stained glass $ASI$ containing the central monopole has a null monopole term and the whole system of the vertex and the boundaries must possibly contain an expansion in terms of dipole, quadrupole, etc. i.e., $V(r) \sim 1/r^{2} + 1/r^{3}+...$.  Therefore, we infer that it is impossible to create an excited charged vertex for an isolated open-edged square-ASI vertex with $z=4$. Since there is no edge spins to compensate the charge at the vertex, only chargeless spin ice state of type-I or type-II can be stabilised with external magnetic field.

\begin{figure}
\includegraphics[width=1\linewidth]{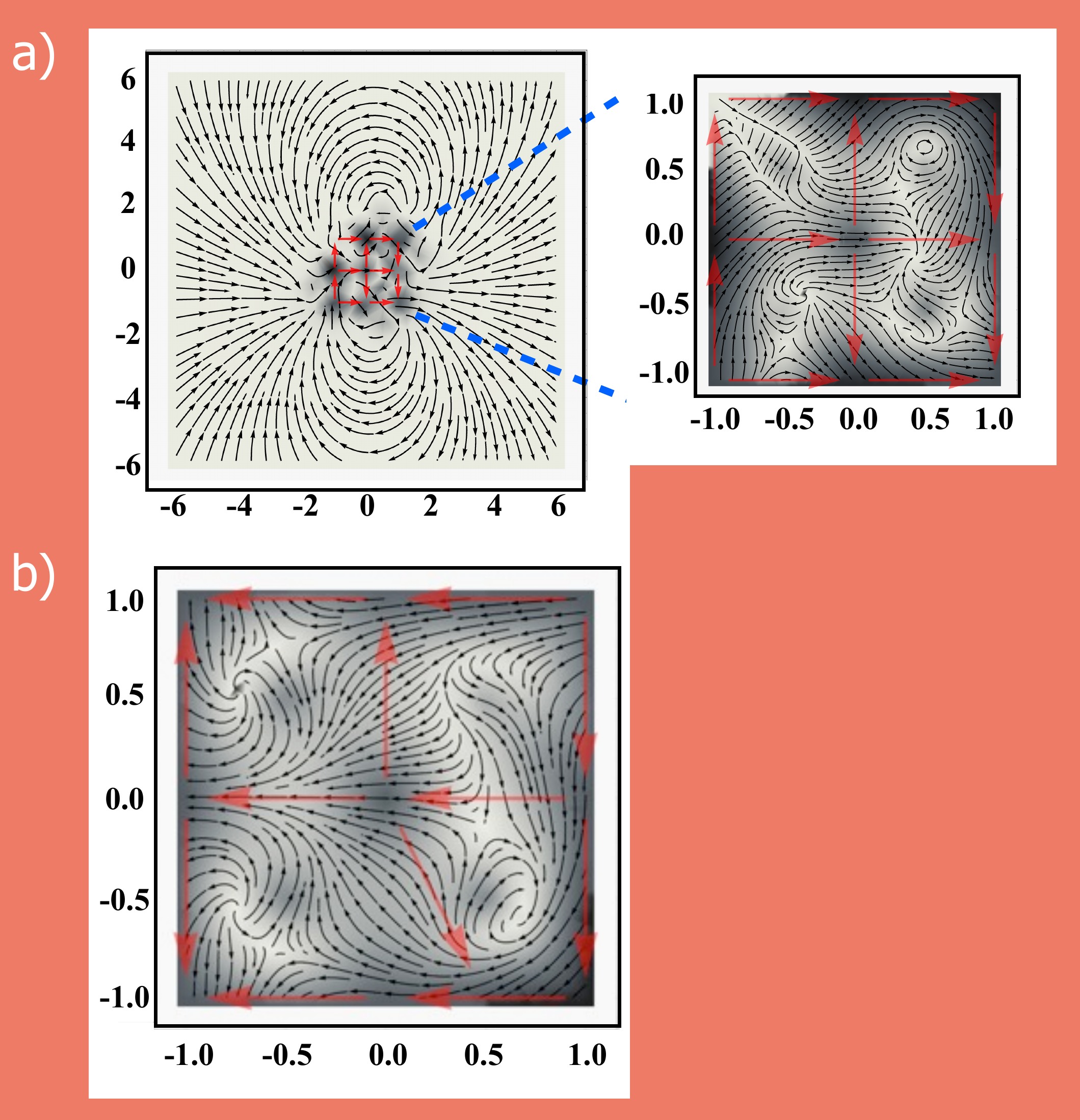}
      \caption{\label{Fig4} \textbf{Magnetic field lines for stained glass (a) and deformed stain glass window (b) samples.} (a) shows the far-field image data. The inset exemplifies the zoomed near-field data. (b) shows near field data of the deformed window. The length scales for both figures are in units of the lattice constant of the artificial lattice.}
\end{figure}

To gain insight into the exact behavior of magnetic field lines, particularly at the monopole states, we performed calculations (see methods) to determine the magnetic field lines for such charged vertices in our stained glass samples. Figure\,\ref{Fig4} shows the field lines for the two samples. Detailed analysis of these results show that the system as a whole resembles the behavior of a magnetic dipole ($\vec{\nabla}\cdot \vec{B}=0$), however, the emergent monopole like behavior is observed as a local effect at the central vertex where $\vec{\nabla}\cdot \vec{B}\neq0$ (Fig.\,\ref{Fig4}a). This is consistent with our analysis of the magnetic charges where we observe the net zero charge for the entire system which is conserved at the emergent monopole state. Similar behavior of local field lines is observed for the deformed window at the monopole state (Fig.\,\ref{Fig4}b). Except for the differences in the local field lines near the edges, the overall behavior exhibited by the deformed and undeformed samples are expectedly similar. These observations of charge neutrality at the monopole states are further confirmed by the Monte Carlo analysis of a broken stained glass window where the 50\% of the edge islands are removed from the undeformed structure. 

The isolated emergent monopole that we observe in these finite-size ASI systems is a classical entity and does not obviously solve Dirac's quantum mechanical problem. However, our work shows for the first time how charges in a finite-size square ASI system are formed at the central vertex with redistribution of charges at the edges so that the charge neutrality is maintained at the single vertex level. 
The results have significant theoretical ramifications. The demonstration of the controlled stabilization of \textit{stringless} isolated emergent monopoles is a significant result particularly since Dirac or Nambu type energetic strings are currently under debate for these systems. Moreover, these results warrant revisit of the concept of the charge neutrality in large arrays of the ASI systems with randomly created monopole-antimonople pairs. In practical terms, our results are important for multiple applications. Plane wave electrons traversing through isolated monopole field acquire azimuthal component due to Aharonov-Bohm effect thereby creating vortex beams of electrons. Interaction of such vortex beams of electrons with matter has potential to lead to multiple applications~\cite{LloydRMP2017}. Our results, thus, pave the way for research on creating electron vortex beams with emergnet monopoles in a controlled way. In addition to the applications in electron vortex beam based research and technology, our results also have a high potential as an involved magnetic source for investigating the amplitudes of scattering of a Dirac particle with charge $Ze$ at relativistic~\cite{Kazama1977} or nonrelativistic ~\cite{BecheNatPhys2014} velocities by an infinitely heavy magnetic monopole (fixed monopole field). Our results will enable performing experiments to test the prediction of helicity flip (100\% polarized with helicity +1 for 180$^{\circ}$ scattered Dirac electrons for $ZeC_m<0$ and helicity -1 for  $ZeC_m>0$) of Dirac's particle scattered from fixed magnetic monopoles~\cite{Kazama1977}.

\end{document}